\documentclass[11pt,english]{extarticle}
\usepackage{listings}
\usepackage{amssymb}
\usepackage{amsfonts}
\usepackage{amsmath}
\usepackage{bm}
\usepackage{tikz}
\usetikzlibrary{matrix,shapes,arrows,fit}
\usepackage{booktabs}
\usepackage{float}
\usepackage{color}
\usepackage[title,titletoc,toc]{appendix}
\usepackage[nohead]{geometry}
\usepackage[bottom]{footmisc}
\usepackage{endnotes}
\usepackage{graphicx}
\usepackage[center,bf]{caption}
\usepackage{morefloats}
\usepackage{epstopdf}
\usepackage{rotating}
\usepackage{enumerate}
\usepackage{bbm}
\usepackage{tikz}
\usepackage{makecell}

\usetikzlibrary{matrix,shapes,arrows,fit,tikzmark}
\tikzset{   
        every picture/.style={remember picture,baseline},
        every node/.style={anchor=base,align=center,outer sep=1.5pt},
        every path/.style={thick},
        }

\tikzstyle{every picture}+=[remember picture] 
\tikzstyle{mybox} =[draw=black, very thick, rectangle, inner sep=10pt, inner ysep=20pt]
\tikzstyle{fancytitle} =[draw=black,fill=red, text=white]
\numberwithin{equation}{section}
\usepackage{fancyhdr}
\usepackage{longtable}
\usepackage{import}
\usepackage{natbib}
\usepackage{microtype}

\usepackage{multirow}
\usepackage{adjustbox}

\usepackage{threeparttable}

\usepackage{verbatim} 
\usepackage{setspace}
\usepackage{pdflscape}
\usepackage{subcaption}
\usetikzlibrary{arrows,calc}
\usepackage{setspace}
\usepackage{ragged2e}
\usepackage{array}

\usepackage{silence}
\usepackage{changepage}

\newcolumntype{L}[1]{>{\raggedright\let\newline\\\arraybackslash\hspace{-2pt}}m{#1}}
\newcolumntype{C}[1]{>{\centering\let\newline\\\arraybackslash\hspace{-2pt}}m{#1}}
\newcolumntype{R}[1]{>{\raggedleft\arraybackslash}p{#1}}

\newlength{\wideitemsep}
\let\olditem\item
\renewcommand{\item}{\setlength{\itemsep}{7pt}\olditem}

\newcommand{\sym}[1]{{#1}}

\long\def\/*#1*/{}

\usepackage{titling}
\renewenvironment{abstract}
{\small
	\begin{center}
		\bfseries \abstractname\vspace{0.0em}\vspace{0pt}
	\end{center}
	\list{}{%
		\setlength{\leftmargin}{6mm}
		\setlength{\rightmargin}{\leftmargin}%
	}%
	\item\relax}
{\endlist}

\usepackage[T1]{fontenc}
\usepackage[latin9]{inputenc}
\usepackage{geometry}
\usepackage{amsthm}
\usepackage{amssymb}

\makeatletter
\theoremstyle{plain}

\theoremstyle{plain}

\ifx\proof\undefined
\newenvironment{proof}[1][\protect\proofname]{\par
	\normalfont\topsep6\p@\@plus6\p@\relax
	\trivlist
	\itemindent\parindent
	\item[\hskip\labelsep\scshape #1]\ignorespaces
}{%
	\endtrivlist\@endpefalse
}
\providecommand{\proofname}{Proof}
\fi
\theoremstyle{plain}

\makeatother

\usepackage{babel}
\providecommand{\corollaryname}{Corollary}
\providecommand{\propositionname}{Proposition}
\providecommand{\theoremname}{Theorem}

\usepackage{xr-hyper}
\usepackage[colorlinks=true,urlcolor=blue, linkcolor=blue, citecolor=blue]{hyperref}
\hypersetup{
filecolor=black
}
\makeatletter
\newcommand*{\addFileDependency}[1]{
  \typeout{(#1)}
  \@addtofilelist{#1}
  \IfFileExists{#1}{}{\typeout{No file #1.}}
}
\makeatother

\begin{document}


\title{\textbf{\large{Democratic Favor Channel}}\thanks{I thank Bryan Caplan, Thiemo Fetzer, John Gerring, Robin Hanson, and Elias Papaioannou. Of course, all errors are mine.}}


\author{\normalsize\textsc{Ziho Park}\thanks{National Taiwan University,  \href{mailto:X@X}{zihopark@ntu.edu.tw}}  \\ \vspace{-5mm} 
}


\date{{\normalsize{}This version: August 5, 2024 \endgraf 
\vspace{0.5mm}
}}
\maketitle

\begingroup
\thispagestyle{empty}
\endgroup   


\begin{abstract}

A large body of literature in economics and political science examines the impact of democracy and political freedoms on various outcomes using cross-country comparisons. This paper explores the possibility that any positive impact of democracy observed in these studies might be attributed to powerful democratic nations, their allies, and international organizations treating democracies more favorably than nondemocracies, a concept I refer to as \emph{democratic favor channel}. Firstly, after I control for being targeted by sanctions from G7 or the United Nations and having military confrontations and cooperation with the West, most of the positive effects of democracy on growth in cross-country panel regressions become insignificant or negatively significant. Secondly, using the same empirical specification as this literature for demonstrating intermediating forces, I show that getting sanctioned, militarily attacked, and not having defense cooperation with the West are plausible channels through which democracy causes growth. Lastly, in the pre-Soviet-collapse period, which coincides with the time when democracy promotion was less often used as a justification for sanctions, the impact of democracy on GDP per capita is already weak or negative without any additional controls, and it becomes further negative once democratic favor is controlled. These findings support the democratic favor channel and challenge the idea that the institutional qualities of democracy per se lead to desirable outcomes. The critique provided in this paper applies to the broader comparative institutions literature in social sciences and political philosophy. \\

Note: This paper is about proposing a new key mechanism that decades-old literature with hundreds of papers largely ignored despite the potential that it can reverse their qualitative conclusions, but it is not just about checking the robustness of specific papers.

\vspace{4mm}

\textbf{\textit{Keywords}}: Institution, Democracy, Autocracy, Sanction
\end{abstract}

\newpage{}
\setcounter{page}{1}

\newpage

\section{Introduction}
\label{section:introduction}

There is a large and important literature on the effect of democracy on various outcomes (e.g, 600 papers between 2000 and 2021 reviewed in \citealt{gerring2022does}), in particular on economic growth (e.g., \citealt{przeworski1995political}, \citealt{barro1996democracy}, \citealt{tavares2001democracy}, \citealt{rodrik2005democratic}, \citealt{persson2006democracy}, \citealt{papaioannou2008democratisation}, \citealt{acemoglu2019democracy}). The works in this decades-old literature generally use cross-country panel data of some outcome variables and democracy indicators. Importantly, the literature interprets the coefficient on democracy as reflecting the institutional qualities of democracy. In this paper, I challenge this interpretation prevalent in this literature by arguing that the positive coefficient may mostly be because powerful democratic nations and international organizations, such as the US and the United Nations, treat democracies more favorably than nondemocracies.

The key institutional knowledge that motivates my critique is the fact that the foreign policies of powerful Western nations and the United Nations often hinge on the ideology that it is their moral obligation and military and economic interests to democratize authoritarian regimes. The argument for exporting democracy frequently surfaces in Western foreign policy discussions. In Congress, government announcements and debates of politicians, policymakers, media, NGO pressure groups, and academics that influence foreign policy, the idea of allying with democracies and confronting autocrats almost always appears. Oftentimes, their argument rests on the morality of democracy. Furthermore, aside from moral and sometimes religious motivation to export democracy, there is a widespread notion that the democratization of foreign countries is in the best interests of Western countries. Among the beliefs in line with this idea is the democratic peace theory, which is among the most popular theories of war and peace, which argues that democratic nations do not engage in wars with each other (e.g., \citealt{maoz1993normative}).\footnote{The argument in this paper does not necessitate that the democratic peace theory is right or wrong. The observation that is sufficient to motivate my critique is that a sizeable fraction, if not most, of those who influence the foreign policies of Western nations or the United Nations believe in theories that justify the propagation of democracy, among them the democratic peace theory.} Note that this theory does not assert that democracies do not engage in wars with autocracies. It asserts that a democracy rarely fights another democracy. Therefore, for the democratic Western policymakers who believe in this theory, it is in their national security interest to democratize foreign countries. 

Consequently, hegemonic democratic countries and the United Nations often treat democracies more favorably than nondemocracies through sanctions and military actions motivated by moral, economic, and security arguments. \cite{felbermayr2020global} have shown that a significant share of sanctions was imposed explicitly to promote democracy and human rights. Also, there are numerous historical examples where the West favored democracies when they decided the level of military confrontation and cooperation with foreign countries. For instance, Bill Clinton embarked on Operation Uphold Democracy in 1994, where the US military forces intervened to remove the military regime in Haiti and reinstated President Aristide, Haiti's first democratically elected president. Moreover, even in cases where the West did not intend to cut military ties with a foreign country, the West's ideological tendency to condemn foreign autocratic oppression eventually led to the dissolution of military cooperation. For example, the US government condemned the Andijan massacre committed by the Uzbek government, resulting in President Islam Karimov ordering the closing of the United States air base in Uzbekistan.

My critique of the literature is that it largely ignores this key channel through which democracy might affect various outcomes, most notably economic growth. Performances of democracies will benefit from pro-democracy favors from the West and the United Nations, and the literature that ignores this channel will provide a faulty interpretation. I will call this idea the \emph{democratic favor channel}. Despite the well-established notion that Western democracies and the United Nations impose sanctions and military operations in a way that favors democracies, the hundreds of existing cross-country works ascribe the differential outcomes of democracies versus non-democracies to their innate institutional qualities, disregarding the potential effects of foreign favors. This is erroneous for the same reason that any study that ascribes the average outcome differences between firms with different corporate governance structures to innate governance qualities while disregarding government policies that benefit certain types of corporate governance is erroneous.

To empirically show that this is a significant concern, I control for sanctions and military confrontation and cooperation in existing works of literature. In the current version of the paper, although the main message of this paper is not about the specific paper but about the broad problem shared by hundreds of works in the literature, I revisit \cite{acemoglu2019democracy} and find support for the democratic favor channel. That is, I analyze the country panel data and regress GDP per capita on democracy and lags of GDP per capita. I show that, once I control for being targeted by sanctions from the West or the United Nations, most of their results turn insignificant or negatively significant. This reversal of the estimated impact of democracy becomes even sharper once I further control military confrontations and cooperation with the West. Moreover, in the pre-Soviet-collapse period studied by \cite{barro1996democracy} which coincides with the time when democracy promotion was less often used as a justification for sanctions by Western nations, the negative impact of democracy is even stronger.

To further empirically demonstrate that democratic favors are the key channel through which democracy "causes" growth, I regress various foreign policies of the West and the UN on democracy, lags of GDP per capita, and lags of the outcome variables. As in \cite{acemoglu2019democracy}, the lags of GDP per capita remove the potential reverse causality so that the empirical model captures the intermediating effect of democratic favors. I find that, consistent with the democratic favor channel, democracy reduces the likelihood of getting sanctioned, militarily attacked, and denied defense cooperation with the West. These patterns are consistent with the idea that democracy "caused" growth because powerful Western countries and the UN favored democracies, not because of institutional qualities. 

The critique provided in this paper applies to the broader comparative institutions literature in social sciences and political philosophy. Not only the large and old empirical literature on the cross-country panel analyses on the effect of democracy but also more non-quantitative works on the role of democracy must also take the democratic favor channel seriously.

Moreover, the democratic favor channel is likely to be even stronger in reality than what the empirics suggest in this paper. That is because there are numerous cases where actual sanctions and military confrontation were considered and nearly implemented due to the autocratic nature of the target countries, but were not implemented in the end. For instance, although the Tiananmen Square Massacre inspired the US House of Representatives to vote to revoke normal trade relations with China, it did not pass in the Senate. Yet, although it was not revoked in the end, the increased trade policy uncertainty would have hampered business relations such as offshoring with China,\footnote{\cite{pierce2016surprisingly} show that granting China \emph{permanent} normal trade relations with China, so that the US Congress does not have to vote on the renewal of normal trade relations annually, led to a sharp increase in imports from China to the US. They also show that pre-PNTR trade uncertainty with China, such as that caused by the negative reaction of the US Congress on the Tiananmen Square Massacre, was a serious concern for businesses. This result highlights the significant role of trade policy uncertainty in hampering trade, even without any change in actual tariffs.} but it will not be captured by any of the controls I included in this paper. Moreover, there are cases where the US considered, but did not execute, cutting military ties with foreign countries due to their autocratic governance. For example, President Jimmy Carter contemplated withdrawing US troops from South Korea in the late 1970s as South Korea's authoritarian suppression intensified, but the concerns about the communist expansion in the Korean peninsula prevented it from actually happening. In this case, although the worsening of US-Korea relations concerned businesses at that time, such an economic impact will also not be captured in my analysis. Therefore, again, even the sharp weakening, often reversal, of the effect of democracy in my empirical results is likely to have been even stronger if I could include all kinds of democratic favors exhaustively.

Section \ref{section:data} explains the data that I used in this study. Section \ref{section:empirical_strategy_in_literature} explains the empirical framework in the literature, and what it misses. Section \ref{section:cause_favor} shows that democracy "causes" democratic favors from the West and the UN. Section \ref{section:control_favor} shows that the positive impact of democracy on growth found in the literature substantially weakens or turns significantly negative once sanctions and military operations by G7 countries and the UN are controlled. Section \ref{section:pre1991} finds that the positive impact of democracy on growth is particularly weak or negative in the period 1960-1991, and the democratic favor channel is observed in this period as well. Section \ref{section:discussion} provides a further discussion on the democratic favor channel. Section \ref{section:conclusion} concludes.

\section{Data and Measurement}
\label{section:data}

\noindent \textbf{Sanctions by the West and the United Nations} \hspace{.3cm} I use the global sanctions database developed by \cite{felbermayr2020global}, which was later extended by \cite{kirikakha2021global} and \cite{syropoulos2024global}. From this data, I obtain country-by-year observations that are sanctioned by each G7 country (US, UK, France, Germany, Canada, Italy, and Japan).  

I collect data on sanctions actually imposed by the United Nations from two sources. One is \cite{biersteker2018targeted} which provides information on UN sanctions from 1991 to 2013, and another is \cite{morgan2014threat}, which covers periods from 1945 to 2005.

I construct two variables. One is an indicator of being sanctioned by the US. As the dominant hegemonic country during the sample period, it turns out that adding this variable significantly harms the target country's economy and removes much of the positive effect of democracy, which is consistent with the democratic favor channel. Another is an indicator of being sanctioned by either one of the other G7 countries or the United Nations. This variable further inflicts a negative effect on the target's economy, and further reduces the positive effect of democracy. \\

\noindent \textbf{Military Confrontation and Cooperation with the West} \hspace{.3cm} I use the Uppsala Conflict Data Program (UCDP)'s External Support Dataset (ESD) (\citealt{meier2023external}) to construct an indicator of warring with the West. On top of that, I manually incorporate Western interventions in civil wars. 

To construct the country panel of significant defense cooperation with major Western nations, I use a few data sources: dyadic data of formal alliances (\citealt{gibler2008international}) obtained from the Correlates of War, country panel of NATO membership, and country panel of having a US military base. I construct an indicator which takes 1 if a country-year has a formal alliance with the US, is a NATO member, or has a US military base. \\

\noindent \textbf{Democracy} \hspace{.3cm} I use democracy indicators developed by various organizations and researchers, such as Freedom House, Polity IV, \cite{cheibub2010democracy}, and \cite{boix2013complete}, \cite{papaioannou2008democratisation}, and \cite{acemoglu2019democracy}. In the tables included in this paper, I use the version used by \cite{acemoglu2019democracy}, but the key finding that taking the democratic favor channel seriously significantly diminishes (or often reverses the sign of) the estimated positive effect of democracy is maintained regardless of whichever democracy indicator I use.

\section{Empirical Framework in the Literature}\label{section:empirical_strategy_in_literature}

Although the large literature of the cross-country panel studies on the impact of democracy largely ignores the democratic favor channel, its analytic framework is useful. Although there are hundreds of studies in this literature, \cite{acemoglu2019democracy} provide several innovations in the specification so I take their empirical model, which takes the following form:

\begin{equation}\label{equa:main_equation_ANRR}
    y_{c t}=\beta D_{c t}+\sum_{j=1}^{p} \gamma_{j} y_{c t-j}+\alpha_{c}+\delta_{t}+\varepsilon_{c t},
\end{equation}

where $y_{c t}$ is the log GDP per capita in country $c$ in year $t$ and $D_{c t}$ is the indicator of democracy in country $c$ in year $t$. The second term on the right-hand-side is $p$ lags of log GDP per capita which controls for the dip in GDP that often preceds democratization. The country fixed effects $\alpha_{c}$ and the year fixed effects $\delta_{t}$ are included. Equation \ref{equa:main_equation_ANRR} is estimated using within-estimator, \cite{arellano1991some}'s GMM estimator, and \cite{hahn2001bias} estimator. Moreover, 2SLS estimates where the democracy indicator is instrumented by regional waves of democracy are also provided.

Equation \cite{acemoglu2019democracy} studies the impact of democracy on GDP per capita. The specification to study the channel through which democracy affects GDP per capita is as follows:

\begin{equation}\label{equa:mechanism_ANRR}
    m_{c t}=\beta D_{c t}+\sum_{j=1}^{p} \gamma_{j} y_{c t-j}+\sum_{j=1}^{p} \eta_{j} m_{c t-j}+\alpha_{c}+\delta_{t}+\varepsilon_{c t},
\end{equation}

where $m_{c t}$ is a potential mechanism, such as investment in GDP, TFP, and the share of trade in GDP. An obvious challenge in studying the mechanism is that growth can result in higher $m_{c t}$, not the other way around. The second term on the right-hand side, the lags of GDP per capita, helps to tackle this concern.

However, importantly, various versions of specifications that are akin to Equation \ref{equa:main_equation_ANRR} or \ref{equa:mechanism_ANRR} in the literature largely ignore the democratic favor channel, which is the focus of the current paper. To demonstrate the democratic favor mechanism, I will extend Equation \ref{equa:main_equation_ANRR} by controlling for various foreign policies of the West and the UN that often favor democracies, and I will use those policies as $m_{c t}$ in Equation \ref{equa:mechanism_ANRR} to show that Western favors for democracies are intermediating mechanism.

\section{Democracy Does "Cause" Better Treatment of the West and the UN}
\label{section:cause_favor}

The democratic favor channel is based on the idea that modern Western foreign policy treats democracies and autocracies differently when it makes sanctions and military decisions. In fact, there is an abundance of examples, a few of which were mentioned in the Introduction, where the West and the UN publicly and proudly state that they give favors to democracies and punishments to autocracies. More often than not, among their primary stated goals of foreign policy principles tend to include the promotion of democracy. In this section, I provide a formal empirical test for this idea.

To do so, I build upon Equation \ref{equa:mechanism_ANRR} by replacing $m_{c t}$ with $DemFavor_{c t}$ which includes various foreign policies that are often used by the West and the UN to punish autocracies and give favors to democracies:

\begin{equation}\label{equa:mechanism_demfavor}
    DemFavor_{c t}=\beta D_{c t}+\sum_{j=1}^{p} \gamma_{j} y_{c t-j}+\sum_{j=1}^{p} \eta_{j} DemFavor_{c t-j}+\alpha_{c}+\delta_{t}+\varepsilon_{c t},
\end{equation}

Table \ref{Table: PredictFavor} provides the results where I consider the following four variables for $DemFavor_{c t}$: an indicator of getting sanctioned by the US (columns 1 and 2), an indicator of getting sanctioned either by the United Nations or by at least one of non-US G7 countries (columns 3 and 4), an indicator of warring with the West (columns 5 and 6), and an indicator of having significant military cooperation with the West (column 7 and 8). Odd-numbered columns report within estimates, and even-numbered columns report results from 2SLS regressions instrumented by regional waves of democratization.

As we can see, there is a clear pattern of democratic favors. Democracies are much less likely to be targeted by US sanctions and UN or non-US G7 sanctions. They are also much less likely to have a war with the West, but they are more likely to have significant defense cooperation with the West.

\begin{table}[ht!]
\caption{Effect of Democracy with Controls for Democratic Favors} \label{Table: PredictFavor}
\begin{adjustwidth}{-0.0cm}{}
\begin{center}
\setlength{\tabcolsep}{9pt}
	{\fontsize{9}{13} \selectfont 
    
{
\def\sym#1{\ifmmode^{#1}\else\(^{#1}\)\fi}
\begin{tabular}{lcccccccc}
\hline \hline
& \multicolumn{1}{c}{\makecell[c]{US\\sanction}} 
& \multicolumn{1}{c}{\makecell[c]{US\\sanction}} 
& \multicolumn{1}{c}{\makecell[c]{UN or\\other G7\\ sanction}} 
& \multicolumn{1}{c}{\makecell[c]{UN or\\other G7\\ sanction}} 
& \multicolumn{1}{c}{\makecell[c]{Fought\\West}} 
& \multicolumn{1}{c}{\makecell[c]{Fought\\West}} 
& \multicolumn{1}{c}{\makecell[c]{Military\\Coop}} 
& \multicolumn{1}{c}{\makecell[c]{Military\\Coop}} 
\\
& \multicolumn{1}{c}{(1)} 
& \multicolumn{1}{c}{(2)} 
& \multicolumn{1}{c}{(3)}
& \multicolumn{1}{c}{(4)} 
& \multicolumn{1}{c}{(5)} 
& \multicolumn{1}{c}{(6)} 
& \multicolumn{1}{c}{(7)} 
& \multicolumn{1}{c}{(8)} 
\\
\hline
Dem &      -0.063\sym{***}&      -0.164\sym{***}&      -0.063\sym{***}&      -0.172\sym{***}&      -0.030\sym{***}&      -0.037\sym{***}&       0.008\sym{***}&       0.037\sym{***}\\
                    &     (0.011)         &     (0.035)         &     (0.012)         &     (0.039)         &     (0.008)         &     (0.012)         &     (0.003)         &     (0.012)         \\
\hline
\makecell[l]{2SLS} & &$\checkmark$ & & $\checkmark$& & $\checkmark$& & $\checkmark$ \\
$N$                     &        6393         &        6366         &        6393         &        6366         &        6393         &        6366         &        6393         &        6366         \\

\hline\hline

\end{tabular}

}
 
        }
\end{center}
\end{adjustwidth}
\footnotesize{
\begin{justify}
\textit{Notes.} This table reports the regression outputs of Equation \ref{equa:mechanism_demfavor} where the dependent variable is various foreign policies of the West and the United Nations that often favor democracies. Columns 1, 3, 5, and 7 provide OLS results. Columns 2, 4, 6, and 8 provide results from 2SLS regressions instrumented by regional waves of democratization. Robust standard errors clustered by country are provided in parentheses. *, **, and *** denote significance at the 10\%, 5\%, and 1\% levels, respectively.
\end{justify}
}
\end{table}

\newpage

\section{Controlling for Democratic Favors}
\label{section:control_favor}

In this section, I additionally control for $DemFavor_{c t}$ to Equation \ref{equa:mechanism_ANRR}:

\begin{equation}\label{equa:main_equation_with_demfavor}
    y_{c t}=\beta_0 D_{c t}+\beta_1 DemFavor_{c t}+\sum_{j=1}^{p} \gamma_{j} y_{c t-j}+\alpha_{c}+\delta_{t}+\varepsilon_{c t},
\end{equation}

In this section, I only show the results from 2SLS regressions although the key idea that controlling for the democracy-favored foreign policies erases or reverses much of the positive impact of democracy is maintained when I instead use within estimator, GMM, or \cite{hahn2001bias} estimator.

Table \ref{Table: Control_favor_19602010} reports the result. Each column provides a slightly different specification. Column 1 uses a one-year lag of the regional wave of democratization for the instrument, and all other columns include four years of lag.

Panel A contains the result without adding $DemFavor_{c t}$, which is exactly the same as \cite{acemoglu2019democracy}. The effect of democracy is mostly positive regardless of specifications albeit insignificantly in a few columns. However, once I control for being sanctioned by the US in Panel B, the estimated effect changes signs in the first three columns, and loses economic and statistical significance in all other columns. As expected, the coefficients on US sanctions are significantly and robustly negative. Panel C additionally controls for all three other elements of $DemFavor_{c t}$. Now most coefficients are negative, and it is statistically significantly negative in column 1. The negative magnitude in column 1 of Panel C is greater than the positive magnitude in column 1 of Panel A. In other words, if the West and the UN were not favoring foreign democracies over autocracies, democracy might have led to a lower GDP per capita. Again, the coefficients on $DemFavor_{c t}$ still remain consistent with intuition even if multiple variables were included simultaneously.

\begin{table}[ht!]
\caption{Effect of Democracy with Controls for Democratic Favors} \label{Table: Control_favor_19602010}
\begin{adjustwidth}{-0.0cm}{}
\begin{center}
\setlength{\tabcolsep}{3pt}
	{\fontsize{9}{13} \selectfont 
    {
\def\sym#1{\ifmmode^{#1}\else\(^{#1}\)\fi}
\begin{tabular}{l*{9}{c}}
\hline\hline
\multirow{3}{10mm}{} 
                     & \multicolumn{3}{c}{} & \multicolumn{5}{c}{Covariates Included} \\
                     \cline{4-10}
                     
    & \multicolumn{1}{c}{} 
    & \multicolumn{1}{c}{} 
    & \multicolumn{1}{c}{\makecell[c]{GDP in 1960 \\ Quintiles\\ $\times$ Year\\ Effects}} 
    & \multicolumn{1}{c}{\makecell[c]{Soviet \\ Dummies}} 
    & \multicolumn{1}{c}{\makecell[c]{Regional \\ Trends}} 
    & \multicolumn{1}{c}{\makecell[c]{Regional \\ GDP and \\Trade}} 
    & \multicolumn{1}{c}{\makecell[c]{Regional \\ Unrest \\GDP  and \\Trade}} 
    & \multicolumn{1}{c}{\makecell[c]{Spatial \\Lag of \\GDP}} 
    & \multicolumn{1}{c}{\makecell[c]{Lags of \\ Democratic \\ Structure}} \\
                    &\multicolumn{1}{c}{(1)}&\multicolumn{1}{c}{(2)}&\multicolumn{1}{c}{(3)}&\multicolumn{1}{c}{(4)}&\multicolumn{1}{c}{(5)}&\multicolumn{1}{c}{(6)}&\multicolumn{1}{c}{(7)}&\multicolumn{1}{c}{(8)}&\multicolumn{1}{c}{(9)}\\
                    
\hline
\multicolumn{1}{c}{} & \multicolumn{9}{c}{\textbf{Panel A}} \\
\cline{2-10}
\makecell[c]{Dem}  &   0.966\sym{*}  &       1.149\sym{**} &       1.125         &       1.292\sym{**} &       1.697\sym{*}  &       1.817\sym{***}&       1.107\sym{*}  &       1.335\sym{**} &       1.361         \\
                    &     (0.558)         &     (0.554)         &     (0.689)         &     (0.651)         &     (0.885)         &     (0.663)         &     (0.656)         &     (0.536)         &     (0.895)         \\

\noalign{\vspace{1em}} 
\makecell[c]{N}      &        6312         &        6309         &        5496         &        6309         &        6309         &        6309         &        6309         &        6181         &        6009         \\

\hline
\multicolumn{1}{c}{} & \multicolumn{9}{c}{\textbf{Panel B}} \\
\cline{2-10}
\makecell[c]{Dem}&      -0.186         &      -0.054         &      -0.240         &       0.280         &       0.948         &       0.906         &       0.353         &       0.104         &       0.162         \\
                    &     (0.677)         &     (0.664)         &     (0.879)         &     (0.724)         &     (0.945)         &     (0.741)         &     (0.749)         &     (0.630)         &     (0.872)         \\

\noalign{\vspace{1em}} 
\makecell[c]{US}&      -1.906\sym{***}&      -1.874\sym{***}&      -1.681\sym{***}&      -1.981\sym{***}&      -1.666\sym{***}&      -1.652\sym{***}&      -1.586\sym{***}&      -1.845\sym{***}&      -1.776\sym{***}\\ \makecell[c]{Sanction}
                    &     (0.339)         &     (0.337)         &     (0.365)         &     (0.358)         &     (0.351)         &     (0.344)         &     (0.349)         &     (0.330)         &     (0.342)         \\

\noalign{\vspace{1em}}                    
\makecell[c]{N}        &        6312         &        6309         &        5496         &        6309         &        6309         &        6309         &        6309         &        6181         &        6009         \\
\hline
\multicolumn{1}{c}{} & \multicolumn{9}{c}{\textbf{Panel C}} \\
\cline{2-10}
\makecell[c]{Dem}&      -1.291\sym{*}  &      -1.144         &      -1.030         &      -0.763         &       0.368         &       0.175         &      -0.360         &      -0.946         &      -0.962         \\
                    &     (0.774)         &     (0.758)         &     (1.017)         &     (0.774)         &     (0.979)         &     (0.795)         &     (0.816)         &     (0.719)         &     (1.031)         \\
       
\noalign{\vspace{1em}} 
\makecell[c]{US}   &      -1.463\sym{***}&      -1.441\sym{***}&      -1.256\sym{***}&      -1.562\sym{***}&      -1.251\sym{***}&      -1.233\sym{***}&      -1.138\sym{***}&      -1.419\sym{***}&      -1.357\sym{***}\\ \makecell[c]{
sanction}
                    &     (0.322)         &     (0.320)         &     (0.346)         &     (0.328)         &     (0.328)         &     (0.318)         &     (0.321)         &     (0.313)         &     (0.334)         \\
     
\noalign{\vspace{1em}} 
\makecell[c]{UN or
other}  &      -1.127\sym{***}&      -1.112\sym{***}&      -1.025\sym{***}&      -1.208\sym{***}&      -0.940\sym{***}&      -1.045\sym{***}&      -1.129\sym{***}&      -1.080\sym{***}&      -1.040\sym{***}\\  \makecell[c]{G7
sanction} 
                    &     (0.265)         &     (0.265)         &     (0.288)         &     (0.282)         &     (0.257)         &     (0.257)         &     (0.258)         &     (0.270)         &     (0.304)         \\

\noalign{\vspace{1em}} 
\makecell[c]{ Fought}      &      -1.101         &      -1.066         &      -1.215         &      -2.393\sym{**} &      -0.915         &      -0.845         &      -0.831         &      -1.049         &      -1.600         \\ \makecell[c]{West}
                    &     (0.980)         &     (0.981)         &     (0.893)         &     (1.010)         &     (1.000)         &     (0.971)         &     (0.967)         &     (0.983)         &     (1.051)         \\

\noalign{\vspace{1em}} 
\makecell[c]{Military
}         &       1.319\sym{*}  &       1.293\sym{*}  &       0.568         &       1.931\sym{**} &       1.349         &       0.929         &       0.970         &       1.141         &       0.669         \\
 \makecell[c]{Coop
}                    &     (0.787)         &     (0.784)         &     (0.937)         &     (0.753)         &     (0.838)         &     (0.784)         &     (0.793)         &     (0.874)         &     (1.266)         \\

\noalign{\vspace{1em}}                    
\makecell[c]{N}          &        6312         &        6309         &        5496         &        6309         &        6309         &        6309         &        6309         &        6181         &        6009         \\
             
\hline
\hline
\end{tabular}
}
 
        }
\end{center}
\end{adjustwidth}
\footnotesize{
\begin{justify}
\textit{Notes.} This table reports the regression results of Equation \ref{equa:main_equation_with_demfavor}. Column 1 uses a one-year lag of the regional wave of democratization for the instrument, and all other columns include four years of lag. Panel A contains the result without adding $DemFavor_{c t}$, which is exactly the same as \cite{acemoglu2019democracy}. Panel B controls for being sanctioned by the US. Panel C additionally controls for being sanctioned by the United Nations or by one of the non-US G7 countries, fighting a war with the West, and having significant military cooperation with the West. Robust standard errors clustered by country are provided in parentheses. *, **, and *** denote significance at the 10\%, 5\%, and 1\% levels, respectively.
\end{justify}
}
\end{table}

\newpage

\section{Restricting to Pre-1991 Period}
\label{section:pre1991}

One might wonder whether the main critique I provided applies to specific periods only. Nonetheless, the nature of autocracy-targeting sanctions, more broadly the degree to which Western nations try to export democracy, has changed over time. In this section, I specifically study the periods from 1960 to 1991. Since the years 1991 and 1992 observed significant turning points in world history, such as the collapse of the Soviet Union and the Treaty on the European Union, the \emph{democratic favor channel} might play out differently for the periods before 1991. Moreover, the period up to 1991 is the era studied in important works such as \cite{barro1996democracy}, which concluded that democracy does not cause economic growth. Moreover, importantly, according to \cite{felbermayr2020global}, the share of sanctions that were explicitly for the purpose of human rights and democracy increased substantially over time. This can be indicative of a change in the principle underlying Western foreign policy toward giving more relative favors for democracies over time. Then, if the positive effect of democracy on growth found in the literature were largely due to democratic favors as I argue, then we will see a weaker positive relationship between democracy and growth in the earlier period.

Table \ref{Table: Control_favor_19601991} restricts the sample to years before 1991. Panel A simply restricts the sample of the baseline 2SLS regressions in \cite{acemoglu2019democracy} to years up to 1991 while not including any new controls. As is evident, this mere restriction of the sample period results in the reversal of the coefficient from positive to negative. That is, there is no positive impact of democracy on growth for the period from 1960 to 1991, the Cold War era.

Panels B and C provide my main results that support the \emph{democratic favor channel}. In Panel B, by including the indicator of being targeted by sanctions from the US, the already-negative coefficients in Panel A become much more significant. In other words, if the US were not more likely to sanction autocracies than democracies, democracy would have caused slower growth during the Cold War era. The negative significance further becomes sharper if I additionally include sanctions from non-US G7 countries, military confrontations and cooperation with the West, and EU membership. In both Panels B and C, the coefficients on getting sanctioned are significantly negative, consistent with the economically destructive effect of sanctions. Fighting the West and having a defense cooperation with the West also have signs that suggest that having a good relationship with the West was important for growth during the Cold War era.

\begin{table}[ht!]
\caption{Effect of Democracy with Controls for Democratic Favors (Up to 1991)} \label{Table: Control_favor_19601991}
\begin{adjustwidth}{-0.0cm}{}
\begin{center}
\setlength{\tabcolsep}{3pt}
	{\fontsize{9}{13} \selectfont 
    {
\def\sym#1{\ifmmode^{#1}\else\(^{#1}\)\fi}
\begin{tabular}{l*{9}{c}}
\hline\hline
\multirow{3}{10mm}{} 
                     & \multicolumn{3}{c}{} & \multicolumn{5}{c}{Covariates Included} \\
                     \cline{4-10}
                     
    & \multicolumn{1}{c}{} 
    & \multicolumn{1}{c}{} 
    & \multicolumn{1}{c}{\makecell[c]{GDP in 1960 \\ Quintiles\\ $\times$ Year\\ Effects}} 
    & \multicolumn{1}{c}{\makecell[c]{Soviet \\ Dummies}} 
    & \multicolumn{1}{c}{\makecell[c]{Regional \\ Trends}} 
    & \multicolumn{1}{c}{\makecell[c]{Regional \\ GDP and \\Trade}} 
    & \multicolumn{1}{c}{\makecell[c]{Regional \\ Unrest \\GDP  and \\Trade}} 
    & \multicolumn{1}{c}{\makecell[c]{Spatial \\Lag of \\GDP}} 
    & \multicolumn{1}{c}{\makecell[c]{Lags of \\ Democratic \\ Structure}} \\
                    &\multicolumn{1}{c}{(1)}&\multicolumn{1}{c}{(2)}&\multicolumn{1}{c}{(3)}&\multicolumn{1}{c}{(4)}&\multicolumn{1}{c}{(5)}&\multicolumn{1}{c}{(6)}&\multicolumn{1}{c}{(7)}&\multicolumn{1}{c}{(8)}&\multicolumn{1}{c}{(9)}\\
                    
\hline
\multicolumn{1}{c}{} & \multicolumn{9}{c}{\textbf{Panel A}} \\
\cline{2-10}
\makecell[c]{Dem} &      -1.202         &      -1.138         &      -0.066         &      -1.249         &      -2.330\sym{**} &      -1.244         &      -0.992         &      -0.596         &      -0.666         \\
                    &     (1.010)         &     (1.017)         &     (0.968)         &     (1.033)         &     (1.185)         &     (0.969)         &     (0.988)         &     (0.834)         &     (0.908)         \\
\noalign{\vspace{1em}} 
\makecell[c]{N}  &        3111         &        3108         &        2774         &        3108         &        3108         &        3108         &        3108         &        3005         &        2860             \\
\hline
\multicolumn{1}{c}{} & \multicolumn{9}{c}{\textbf{Panel B}} \\
\cline{2-10}
\makecell[c]{Dem} &      -1.902\sym{*}  &      -1.837\sym{*}  &      -0.739         &      -1.958\sym{*}  &      -2.939\sym{**} &      -1.784\sym{*}  &      -1.449         &      -1.277         &      -1.446         \\
                    &     (1.085)         &     (1.083)         &     (0.992)         &     (1.100)         &     (1.275)         &     (1.049)         &     (1.104)         &     (0.883)         &     (0.922)        \\
\noalign{\vspace{1em}}
\makecell[c]{US}  &      -1.991\sym{***}&      -1.977\sym{***}&      -1.566\sym{***}&      -2.011\sym{***}&      -2.154\sym{***}&      -2.078\sym{***}&      -1.930\sym{***}&      -1.912\sym{***}&      -2.033\sym{***}\\
 \makecell[c]{Sanctions}    &     (0.608)         &     (0.613)         &     (0.549)         &     (0.616)         &     (0.670)         &     (0.628)         &     (0.630)         &     (0.595)         &     (0.631)          \\
\noalign{\vspace{1em}}                    
\makecell[c]{N}    &        3111         &        3108         &        2774         &        3108         &        3108         &        3108         &        3108         &        3005         &        2860      \\

\hline
\multicolumn{1}{c}{} & \multicolumn{9}{c}{\textbf{Panel C}} \\
\cline{2-10}
\makecell[c]{Dem}&      -1.913\sym{*}  &      -1.841\sym{*}  &      -0.738         &      -1.942\sym{*}  &      -2.924\sym{**} &      -1.873\sym{*}  &      -1.544         &      -1.317         &      -1.729\sym{*}  \\
                    &     (1.092)         &     (1.077)         &     (1.001)         &     (1.098)         &     (1.274)         &     (1.051)         &     (1.116)         &     (0.859)         &     (0.970)         \\

\noalign{\vspace{1em}} 
\makecell[c]{US}   &      -1.605\sym{***}&      -1.590\sym{***}&      -1.234\sym{**} &      -1.576\sym{***}&      -1.653\sym{***}&      -1.747\sym{***}&      -1.588\sym{***}&      -1.544\sym{***}&      -1.678\sym{***}\\ \makecell[c]{
sanction}
                    &     (0.539)         &     (0.543)         &     (0.498)         &     (0.545)         &     (0.588)         &     (0.547)         &     (0.551)         &     (0.522)         &     (0.563)         \\

\noalign{\vspace{1em}} 
\makecell[c]{UN or
other}  &      -1.552\sym{***}&      -1.546\sym{***}&      -1.261\sym{**} &      -1.537\sym{***}&      -1.597\sym{***}&      -1.335\sym{**} &      -1.493\sym{***}&      -1.597\sym{***}&      -1.524\sym{***}\\
   \makecell[c]{G7
sanction}                  &     (0.562)         &     (0.564)         &     (0.631)         &     (0.568)         &     (0.544)         &     (0.530)         &     (0.532)         &     (0.565)         &     (0.572)         \\

\noalign{\vspace{1em}} 
\makecell[c]{ Fought}     &      -4.668\sym{***}&      -4.656\sym{***}&      -4.163\sym{***}&      -5.546\sym{***}&      -4.543\sym{***}&      -4.602\sym{***}&      -4.535\sym{***}&      -4.566\sym{***}&      -5.278\sym{***}\\ \makecell[c]{West}
                    &     (1.482)         &     (1.487)         &     (1.484)         &     (1.484)         &     (1.517)         &     (1.497)         &     (1.488)         &     (1.521)         &     (1.776)         \\

\noalign{\vspace{1em}} 
\makecell[c]{Military
}        &       2.493\sym{**} &       2.494\sym{**} &       1.165         &       2.429\sym{**} &       2.402\sym{**} &       2.477\sym{**} &       2.099         &       2.997\sym{**} &       4.005\sym{***}\\ \makecell[c]{Coop} 
                    &     (1.117)         &     (1.117)         &     (1.193)         &     (1.095)         &     (1.178)         &     (1.109)         &     (1.280)         &     (1.283)         &     (1.066)         \\

\noalign{\vspace{1em}}                    
\makecell[c]{N}     &        3111         &        3108         &        2774         &        3108         &        3108         &        3108         &        3108         &        3005         &        2860         \\

\hline
\hline
\end{tabular}
}
 
        }
\end{center}
\end{adjustwidth}
\footnotesize{
\begin{justify}
\textit{Notes.} This table reports the regression results of Equation \ref{equa:main_equation_with_demfavor} while restricting the sample to years between 1960 and 1991. Column 1 uses a one-year lag of the regional wave of democratization for the instrument, and all other columns include four years of lag. Panel A contains the result without adding $DemFavor_{c t}$, which is exactly the same as \cite{acemoglu2019democracy}. Panel B controls for being sanctioned by the US. Panel C additionally controls for being sanctioned by the United Nations or by one of the non-US G7 countries, fighting a war with the West, and having significant military cooperation with the West. Robust standard errors clustered by country are provided in parentheses. *, **, and *** denote significance at the 10\%, 5\%, and 1\% levels, respectively.
\end{justify}
}
\end{table}

\newpage

\section{Discussion}
\label{section:discussion}

\noindent \textbf{Accidental democratic favors?} \hspace{.3cm} One might attempt to counter my critique by arguing that the sanctions and military operations might have favored democracies merely accidentally but not intentionally. That is, although the US, other G7 nations, and the UN might claim that their goal is the propagation of democracy, their true intentions would have been something else. That is, the claim would be that the West and the UN would have merely sanctioned foreign countries based on the targets' characteristics that happened to have a positive correlation with autocracy. For instance, during the Cold War era, the US and other Western nations sanctioned the Soviet Union and other communist countries, most (if not all) of which were dictatorial, so that a sanction targeting communists would accidentally target autocrats. 

The most straightforward rebuttal to this counterargument to my critique would be numerous accounts of intentional democratic favors that can be found in the statements of the Western governments, the UN, the Congress, and academia, NGOs, and media that influence foreign policy, including a few examples I included in the Introduction. Such an argument is particularly hard to apply for the sanctions imposed by the United Nations, which does not only reflect the national interest of a particular country.

Moreover, perhaps more importantly, even in the unlikely case that sanctions and military operations merely accidentally favored democracies, that will not weaken my critique of the literature. Such a coincidental democratic favor will still be an important channel through which democracy affects growth not because of its institutional superiority but because of foreign interference, regardless of whether it was intentionally or coincidentally targeting authoritarian regimes. \\

\noindent \textbf{Autocracy itself causing foreign confrontations?} \hspace{.3cm} Another potential counterargument to the democratic favor channel would be that the institutional characteristics of autocracy per se cause confrontations with foreign hegemons, even if those hegemons do not aim to export democracy. Therefore, the argument would imply that we can still say that democratic institutions cause growth. However, this claim will have an unrealistic implication that, even if counterfactually the US were either a dictatorial hegemon or a nation with no foreign policy ideology of democracy promotion, the degree to which autocracies conflict with the US more than democracies do would have been the same as the reality. This is highly unlikely. For instance, the US condemnation of the authoritarian oppression in Uzbekistan led to the removal of the US military base from the former Soviet Union country. There do not appear to be any material gains for the US as a consequence of condemning the Uzbek dictator other than a willingness to do what they believe to be moral or a belief that democracies benefit from the democratization of foreign countries. It is thus highly improbable that the massacre in Uzbekistan would have led to the removal of the US military base if the US did not have a foreign policy in favor of democracy and human rights.

\newpage

\section{Conclusion}
\label{section:conclusion}

The critique in this paper is applicable to the large literature that studies the impact of democracy and political freedoms on various outcomes. When powerful entities are treating different institutions differently, we ought to factor in foreign favor channels to obtain correct implications for comparative institutions. The democratic favor channel should be taken seriously in the large and old literature on the effect of democracy, including both the modern quantitative empirical panel studies and more non-quantitative philosophical works.

\newpage{}

\bibliographystyle{aer}
\bibliography{Reference}

\end{document}